\documentclass[prl,reprint,showpacs,raggedbottom,nofootinbib,superscriptaddress]{revtex4-1}
\usepackage{bm,dcolumn}
\usepackage{amsmath,amsfonts,amssymb,amsthm,amscd} 
\usepackage{graphicx}
\usepackage{verbatim}
\usepackage{lipsum}
\usepackage[section]{placeins}
\usepackage{color}
\usepackage{soul}
\begin{document}
\sloppy
\title{Anomalous broadening in driven dissipative Rydberg systems}
\author{E. A. Goldschmidt}
\affiliation{United States Army Research Laboratory, Adelphi, Maryland 20783 USA}
\author{T. Boulier}
\affiliation{Joint Quantum Institute, National Institute of Standards and Technology and the University of Maryland, Gaithersburg, Maryland 20899 USA}
\author{R. C. Brown}
\altaffiliation[Present address: ]{National Institute of Standards and Technology, Boulder, Colorado 80305 USA}
\affiliation{Joint Quantum Institute, National Institute of Standards and Technology and the University of Maryland, Gaithersburg, Maryland 20899 USA}
\author{S. B. Koller}
\altaffiliation[Present address: ]{Physikalisch Technische Bundesanstalt, 38116 Braunschweig, Germany}
\affiliation{Joint Quantum Institute, National Institute of Standards and Technology and the University of Maryland, Gaithersburg, Maryland 20899 USA}
\author{J. T. Young}
\affiliation{Joint Quantum Institute, National Institute of Standards and Technology and the University of Maryland, Gaithersburg, Maryland 20899 USA}
\author{A. V. Gorshkov}
\affiliation{Joint Quantum Institute, National Institute of Standards and Technology and the University of Maryland, Gaithersburg, Maryland 20899 USA}
\affiliation{Joint Center for Quantum Information and Computer Science, National Institute of Standards and Technology and the University of Maryland, College Park, Maryland 20742 USA}
\author{S. L. Rolston}
\affiliation{Joint Quantum Institute, National Institute of Standards and Technology and the University of Maryland, Gaithersburg, Maryland 20899 USA}
\author{J. V. Porto}
\email{porto@umd.edu}
\affiliation{Joint Quantum Institute, National Institute of Standards and Technology and the University of Maryland, Gaithersburg, Maryland 20899 USA}
\begin{abstract}
We observe interaction-induced broadening of the two-photon $5s$-$18s$ transition in $^{87}$Rb atoms trapped in a 3D optical lattice. The measured linewidth increases by nearly two orders of magnitude with increasing atomic density and excitation strength, with corresponding suppression of resonant scattering and enhancement of off-resonant scattering. We attribute the increased linewidth to resonant dipole-dipole interactions of $18s$ atoms with blackbody induced population in nearby $np$ states. Over a range of initial atomic densities and excitation strengths, the transition width is described by a single function of the steady-state density of Rydberg atoms, and the observed resonant excitation rate corresponds to that of a two-level system with the measured, rather than natural, linewidth. The broadening mechanism observed here is likely to have negative implications for many proposals with coherently interacting Rydberg atoms.
\end{abstract}
\maketitle

Due to their strong, long-range, coherently-controllable interactions, Rydberg atoms have been proposed as a basis for quantum information processing and simulation of many-body physics \cite{weimer2010,Lee2013,Pohl2011, Saffman2010}. Using the coherent dynamics of such highly excited atomic states, however, requires addressing challenges posed by the dense spectrum of Rydberg levels, the detrimental effects of spontaneous emission, and strong interactions. One approach is to operate on timescales much faster than the long Rydberg lifetime, typically microseconds to milliseconds \cite{heidemann07a,barredo2015,bendkowsky2009,urban2009}. Another proposed approach is to off-resonantly couple the ground and Rydberg state, admixing a small amount of the strongly interacting character into the ground state while substantially reducing spontaneous emission \cite{Johnson2010,henkel10a,pupillo2010,honer10a,Glaetzle2012,vanBijnen2015,Glaetzle2015,Bouchoule2002,Lee2013a,Dauphin2012,Glaetzle2014}. This Rydberg-dressed atom approach has been recently demonstrated with pairs of atoms \cite{Jau2015}, but has been difficult to realize in a many-body context \cite{Balewski2014}. A full understanding of the scope and limitations of these proposals requires including the effects of spontaneous decay within the dense energy level structure, which typically cannot be described by a mean-field treatment in interacting gases due to correlated quantum coherent and dissipative effects. 

We study the effect of interactions in a driven, dissipative system of Rydberg atoms in a 3D optical lattice. We observe significant deviation from the expected excitation rates both on and off resonance that cannot be explained by van der Waals interactions or a mean-field treatment of the system. We attribute these effects to  blackbody induced transitions to nearby Rydberg states of opposite parity, which have large, resonant dipole-dipole interactions with the state of interest. These off-diagonal exchange interactions result in complex many-body states of the system. Previous work has explored the impact of similar, controlled, interactions \cite{Anderson1998,Park2011,Gunter2013,Brekke2012}, however the uncontrolled creation of strongly interacting Rydberg levels due to spontaneous or blackbody processes is typically ignored in discussions of coherent Rydberg dynamics. These interactions may significantly modify the parameter regimes available for many-body Rydberg-based systems. In particular, we show that even at low densities of Rydberg atoms, uncontrolled production of atoms in other states significantly modifies the energy levels of the remaining atoms.

\begin{figure}[!h]
\includegraphics[width=0.95\linewidth]{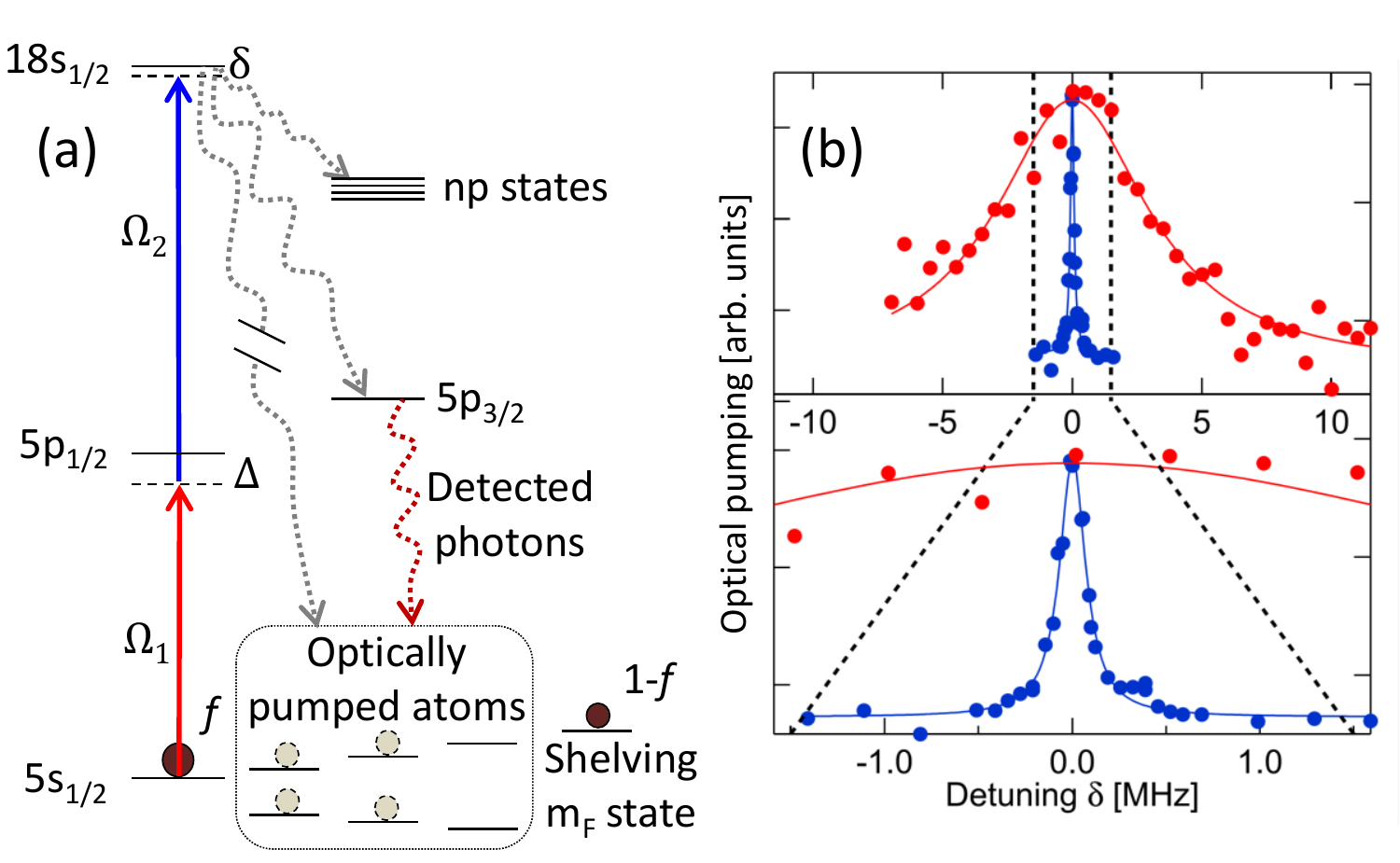}
\caption{a) Level diagram for two-photon excitation to the $18s$ Rydberg level. Fractional density, $f$, in the $\left|F=2,m_F=-2\right\rangle$ ground state is controlled by microwave transfer, remaining atoms are shelved in the non-participating $\left|F=2,m_F=+2\right\rangle$ state (spectrally resolvable due to a $0.3~\rm{mT}$ magnetic field along the optical axis). Decay from the $18s$ state occurs via many channels, including via Rydberg $np$ levels and the $5p_{3/2}$ state. Atoms are optically pumped to the $\left|F=1,2;m_F=-1,0\right\rangle$ ground states. b) Example $18s$ spectra measured as the population in the $m_F=0,\pm1$ states vs two-photon detuning $\delta$. Blue is $\Omega/2\pi$=3~kHz, $f$=0.3 and red is $\Omega/2\pi$=140~kHz, $f$=0.75.} \label{levels}
\end{figure}

We use a state in $^{87}$Rb with principal quantum number $n=18$, and relatively short natural lifetime, $1/\Gamma_0=3.5~\mu$s (including blackbody transitions), to study dissipative effects in a Rydberg system. Atoms in the same Rydberg level interact primarily via the $C_6/r^6$ van der Waals interaction, which for the $18s$ state is repulsive and equals the $18s$ linewidth at 800~nm separation. On the other hand, atoms in states of opposite parity interact via the much larger resonant dipole-dipole interaction, $\propto C_3/r^3$, whose angular dependence allows it to be positive or negative \cite{Beterov2009, Afrousheh2004} and, for $18s$ interacting with $17p$ or $18p$, equals the $18s$ linewidth at $16~\mu$m separation. Blackbody-induced transitions to other Rydberg levels constitute $\gtrsim20~\%$ of the decay. We note that molecular resonances can be ignored due to the low principal quantum number that allows molecule formation only at extremely high densities \cite{Derevianko2015}. 

We excite the $18s_{1/2}$ state using a two-photon transition via the 5p$_{1/2}$ state (Fig. \ref{levels}a), with intermediate state detuning $\Delta/2\pi\approx235$~MHz and independently calibrated single-photon Rabi frequencies, $\Omega_1/2\pi<10$~MHz and $\Omega_2/2\pi\approx7$~MHz \cite{supplemental}. The two excitation lasers are stabilized to the same high-finesse optical cavity with $<10$~kHz linewidth, and are polarized and tuned to couple the ground $\left|5s_{1/2};~F=2,m_F=-2\right\rangle$ hyperfine state to the $\left|18s_{1/2};~F=2,m_F=-2\right\rangle$ state with two-photon detuning $\delta$ and Rabi frequency $\Omega=\Omega_1 \Omega_2/2\Delta$.   

The atomic system consists of a Bose-Einstein condensate of $\approx4\times10^4$ atoms initially in the $\left|F=1,m_F=-1\right\rangle$ ground state, loaded into a 3D optical lattice \cite{supplemental, Sebby-Strabley2006}. The lattice provides a minimum separation of 406~nm and additionally suppresses superradiant Rayleigh scattering on the $5s$-$5p$ transition \cite{Inouye1999}. We control the atomic density available for Rydberg excitation with microwave rapid adiabatic passage that puts a fraction, $f$, of the atoms in the $\left|F=2,m_F=-2\right\rangle$ hyperfine state and shelves all remaining atoms in the non-participating $\left|F=2,m_F=+2\right\rangle$ state. This technique varies the average density, $\rho_g=f \times 57~\mu$m$^{-3}$, without altering the geometry of the cloud. We quantify excitation to the Rydberg state by measuring the population remaining in the initial state (or, equivalently, pumped into the initially empty $m_F=0,\pm1$ states) following excitation. The ground hyperfine populations are separated in time-of-flight with a Stern-Gerlach magnetic field gradient and measured via absorption imaging.

\begin{figure}[!hb]
\includegraphics[width=6cm]{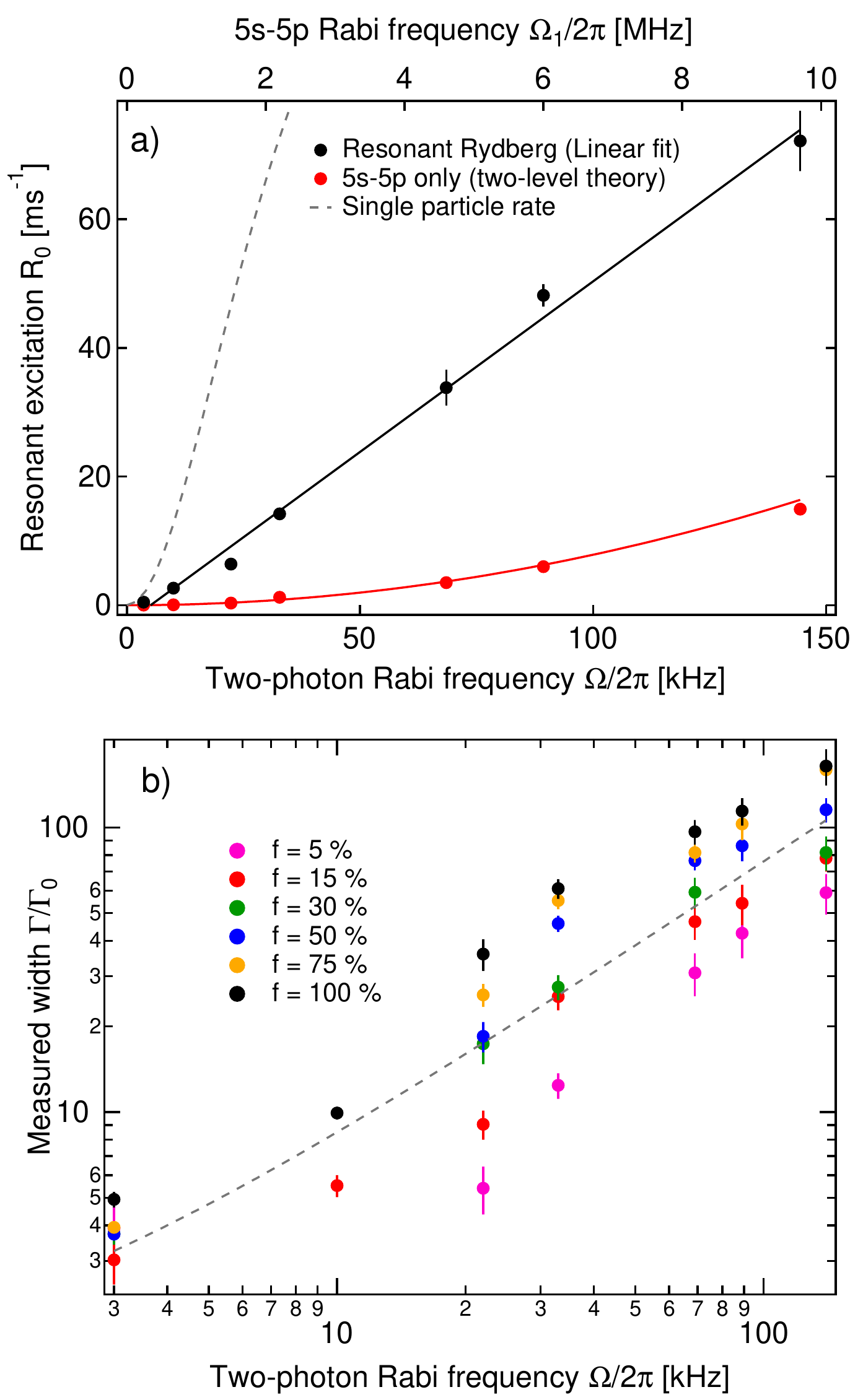}
\caption{a) Measured optical pumping rate for $5s$-$5p$ field only (red) vs $\Omega_1$ (top axis) and two-photon resonant excitation with lower rate subtracted (black) vs $\Omega$ (bottom axis). Black line is a linear fit to $R_0$ and red line is the calculated $5s$-$5p$ rate with no adjustable parameters. Gray dashed line is the expected single particle rate. b) Measured width $\Gamma$ in units of the natural linewidth $\Gamma_0=2\pi\times45~\rm{kHz}$ vs $\Omega$ for different fractional densities. Dashed line is linear scaling. Error bars represent statistical fitting uncertainties.} \label{unscaled}
\end{figure}

We measure the Rydberg excitation rate $R$ (proportional to the optically pumped fraction following a fixed excitation time at least several $18s$ lifetimes, but shorter than the time to depump the initial state) as a function of $\delta$, $\Omega$ and $\rho_g$. Observed lineshapes are symmetric and well-characterized by Lorentzians (Fig.~\ref{levels}b)
\begin{equation}
R = \frac{R_0}{1 + 4 \delta^2/\Gamma^2}.
\end{equation}
We fit a decaying exponential to the population remaining in the initial state as a function of excitation time for two-photon excitation with $\delta=0$ and for the lower $5s$-$5p$ field alone (Fig. \ref{unscaled}a). For each $\Omega$ and $\rho_g$, we extract the resonant excitation rate $R_0$ by subtracting the $5s$-$5p$ optical pumping rate from the measured two-photon rate and scaling by the $45~\%$ fraction that decays to states other than the initial state. The linewidth $\Gamma$ is determined from a Lorentzian fit to the optical pumping as a function of $\delta$ (Fig. \ref{unscaled}b). We observe that $\Gamma$ increases dramatically with both $\Omega$ and $\rho_g$, reaching values as large as $\approx200\Gamma_0$. At small $\Omega$ and $\rho_g$ the narrowest observed linewidth is $\approx3\Gamma_0$ and the residual broadening is attributed to inhomogeneities such as optical trapping light shifts and laser frequency noise. 

Remarkably, $R_0$ is linear in $\Omega$ (with slope that depends on $\rho_g$) and shows no sign of saturation up to $\Omega = 3 \Gamma_0$ (Fig.~\ref{unscaled}a). This behavior is inconsistent with standard single-atom theory and purely inhomogeneous broadening, which predicts faster excitation that depends on $\Omega^2$ for small $\Omega$ and saturates at large $\Omega$ (Fig. \ref{unscaled}a dashed line). The observed $R_0$ corresponds to a single-atom theory assuming the measured $\Gamma$ as the transition linewidth: $R_0\approx\Omega^2/\Gamma$. 

\begin{figure}[!h]
\includegraphics[width=7.5cm]{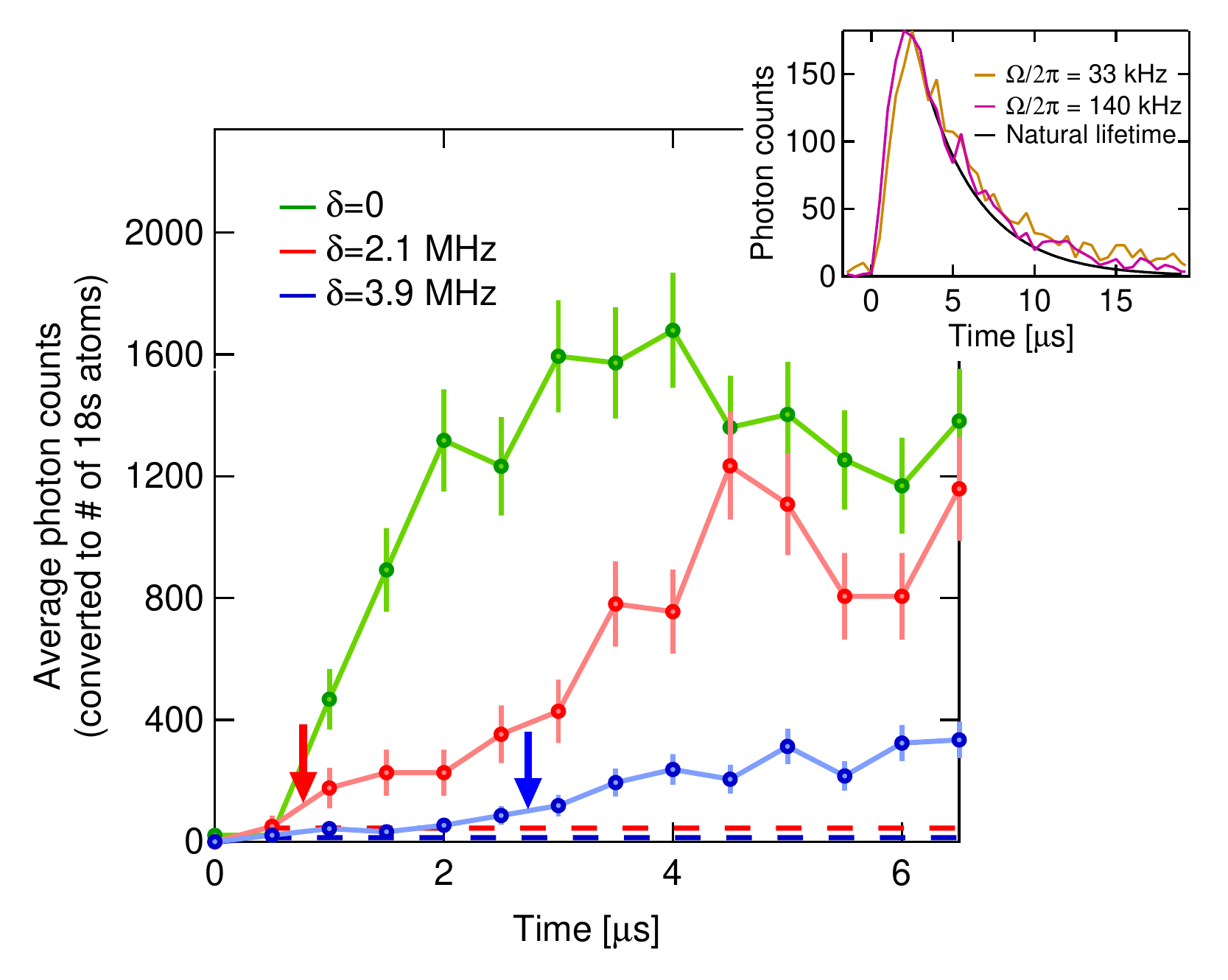}
\caption{Average fluorescence photon counts (converted to number of $18s$ atoms) as the excitation turns on for different two-photon detunings $\delta$ for $\Omega/2\pi$=140~kHz and $f=1$. Dashed lines indicate the number of excited atoms expected for the non-interacting, single particle scattering rates at those detunings. Arrows indicate the expected times to create the first contaminant atom for the detuned cases (see text). Error bars represent shot noise. Inset: Decay following short ($2~\mu$s) pulses for different two-photon Rabi frequencies with natural lifetime plotted for reference. All regimes are consistent with the $18s$ natural lifetime.} \label{lifetime}
\end{figure}

To determine whether the observed broadening corresponds to a concomitant shortening of the $18s$ lifetime, as would broadening due to superradiance \cite{Wang2007}, we collect fluorescence emitted on the $5p_{3/2}$-$5s_{1/2}$ transition (Fig. \ref{levels}a). The fluorescence, which scales with the optical pumping signal and is proportional to the number of $18s$ atoms, is collected by a lens relay system (NA=0.12) with an interference filter to block the $5s$-$5p_{1/2}$ excitation light, detected by a single photon avalanche diode, time-tagged with 21-ns resolution, and summed over many excitation pulses. The observed lifetime, measured as the decay in detected photons after extinguishing the excitation light, is consistent with the 3.5~$\mu$s natural lifetime and independent of $\Omega$ (see Fig. \ref{lifetime} inset and \cite{supplemental} for more information). This result is consistent with previous observations of the suppression of superradiance due to driven dipole interactions \cite{Pellegrino2014}. The confirmation of the natural lifetime, along with the lack of saturation of the optical pumping, rules out superradiance and suggests the broadening is due to rapid dephasing of the optical coherence. In addition, confirmation of the lifetime allows an estimate of the steady-state $18s$ population. 

The steady-state density of $18s$ atoms, under resonant excitation, is the atomic density, $\rho_g$, scaled by the ratio of the excitation rate, $R_0$, to the decay rate, $\Gamma_0$: $\rho_{18s}=\rho_g R_0/\Gamma_0$. The steady-state densities in nearby $np$ states are equal to $\rho_{18s}$ scaled by the ratios of the $18s-np$ transition rates to the $np$ decay rates: $\rho_{np}=\rho_{18s}b_{np}\Gamma_0/\Gamma_{np}=\rho_gR_0b_{np}/\Gamma_{np}$, where $b_{np}$ are the branching ratios from $18s$ to $np$ (dominated by blackbody transitions to $17p$ and $18p$ states), and $\Gamma_{np}$ are the decay rates of the $np$ states (including blackbody transitions). 

We observe $\Gamma$ as large as 8~MHz, inconsistent with the 1.9~MHz van der Waals shift expected at our highest $18s$ densities \cite{supplemental}. In addition, the observed lineshapes are symmetric, inconsistent with the repulsive van der Waals interaction. Also, the broadening depends only on the average density $\rho_g$ and is independent of the microscopic configuration, which we alter by transferring atoms in every other lattice site in 2D to the shelving state~\cite{Lee2007}. The width, lineshape symmetry, and insensitivity to nearest-neighbor spacing are consistent with the larger, longer-range, symmetric, dipole-dipole interaction between states of opposite parity. 

\begin{figure}[!h]
\includegraphics[width=6cm]{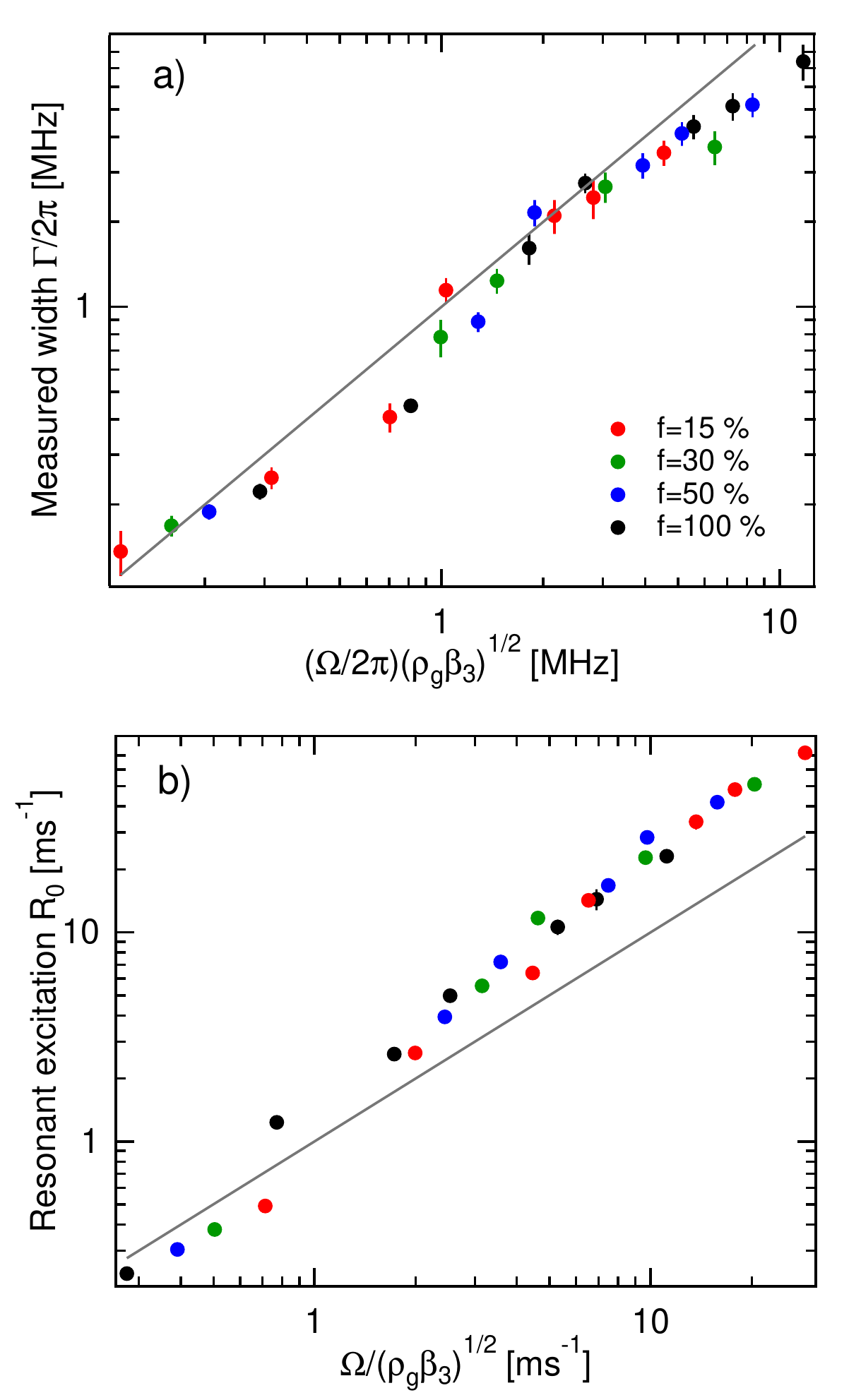}
\caption{(a) Measured width $\Gamma$ vs $\Omega\sqrt{\rho_g\beta_3}$ and (b) resonant excitation rate $R_0$ vs $\Omega/\sqrt{\rho_g\beta_3}$ at different two-photon Rabi frequencies and densities. Linear scalings with unit slope are indicated with solid lines. Error bars represent statistical fitting uncertainties.} \label{collapse}
\end{figure}

For broadening due to dipole interactions, we expect a width of order $\sum\left|C_3^{(np)}\right|\rho_{np}$, where the sum is over the $np$ states, which have different effective interaction strengths and branching ratios. This can be rewritten as $\beta_3\rho_gR_0$ using the expression above for $\rho_{np}$ and defining the quantity $\beta_3=\sum \left|C_3^{(np)}\right| b_{np}/\Gamma_{np}=116~\mu\mathrm{m}^3$ (including the root-mean-squared average of the angular dependence of $C_3$). Combined with the observed relation $R_0\approx\Omega^2/\Gamma$, this provides expressions for $\Gamma$ and $R_0$ in terms of independently controlled variables $\Omega$ and $\rho_g$:
\begin{equation}
\begin{split}
\Gamma\approx\Omega\sqrt{\rho_g\beta_3},\\
R_0\approx\frac{\Omega}{\sqrt{\rho_g\beta_3}}.
\label{vars}
\end{split}
\end{equation} 
$\Gamma$ and $R_0$ are plotted in terms of these expressions in Fig. \ref{collapse}. The data not only collapse to approximately linear curves over two orders of magnitude, but the magnitude is well described by the dipole-dipole energy scale characterized by the independently calculated factor $\beta_3$. (Neither of these features is present for scaling with the van der Waals interaction \cite{supplemental}). This agreement is highly suggestive of a broadening mechanism dominated by dipole interactions with contaminant states. The fluctuating microscopic configuration of $np$ states being populated and decaying leads to dephasing that is not accompanied by either saturated optical pumping or shortening of the lifetime. 

This broadening mechanism requires some initial time to populate the contaminant states. We study the time dynamics of resonant and detuned excitation using the fluorescence on the $5p_{3/2}$-$5s_{1/2}$ transition. Fig. \ref{lifetime} shows the fluorescence, converted into a number of $18s$ atoms, as a function of time for excitation at different detunings with $\Omega/2\pi=140$~kHz and $f=1$. At the two non-zero detunings, the population reaches a significant fraction of the resonantly excited population in a few $\mu$s. This is in stark contrast to the expected single-atom scattering times of $3~\mathrm{ms}$ and $11~\mathrm{ms}$ for these nominally far detuned cases ($\tau_s=4\delta^2/\Gamma_0\Omega^2$ for $\delta\gg\Omega,\Gamma_0$). The faster excitation off resonance leads to observed $18s$ populations larger by a factor of $\gtrsim30$ than expected. For the resonant case, on the other hand, the observed population is smaller than expected from a single-atom picture by a factor of $\gtrsim10$, which cannot be explained by van der Waals interactions alone. The expected non-interacting excitation rates, both on and off resonance, are central to the feasibility of both dressed Rydberg proposals and Rydberg quantum gate implementations. We observe substantial deviations from these expected rates, which must be addressed for a full analysis of any Rydberg system.

This shortening of the off-resonant scattering time may not be a problem in few atom systems such as arrays of microtraps or other 2D systems \cite{Jau2015,Zeiher2015,Xia2015}, but it is likely problematic for implementing Rydberg-dressed atom proposals in large, many-body systems \cite{Johnson2010,henkel10a,pupillo2010,honer10a}. The time until the creation of the first contaminant atom is $\tau=\tau_s/bN_0$, where $b\gtrsim20~\%$ is the branching to contaminant states and $N_0$ is the total number of atoms. Interaction with the first $np$ atom allows excitation of other atoms at a faster, resonant rate, leading to additional $np$ atoms that, in turn, increase the number of atoms resonantly excited, similar to Rydberg aggregation at shorter timescales due to van der Waals shifts \cite{Schempp2014,Urvoy2015,Malossi2014}. The long-range nature of the dipole interaction causes aggregation on length scales comparable to typical experimental system sizes, leading to rapid broadening over the entire ensemble. This simple timescale estimate gives a qualitative understanding of the early time dynamics, and future work to develop a full microscopic model will hopefully provide quantitative descriptions of both the dynamic and steady-state behavior \cite{supplemental}. Finally, the magnitude of the uncontrolled interactions with contaminant atoms is large compared to the interactions in a Rydberg-dressed approach. In particular, a dressed atom's uncontrolled dipole interaction with a contaminant atom is larger than its interaction with another dressed atom for $\rho_{np}>\Omega^2/\delta C_3$ \cite{Balewski2014}, which is quickly exceeded under reasonable experimental conditions \cite{supplemental}.


In conclusion, we report experimental observation of large spectral broadening of a Rydberg transition modifying the scattering rate both on and off resonance. We infer this effect results from the uncontrolled buildup of atoms in nearby Rydberg states. Resonant dipole-dipole interaction with those states causes dephasing and broadens the driven transition. Any single-atom approach to this problem is inherently nonlinear, as the broadening depends on the excited population, leading to distinctly non-Lorentzian lineshapes that contradict our observations. Mean-field approaches fail because the off-diagonal interaction requires single-atom coherences between the driven state and contaminant states, which do not develop under blackbody induced population of the contaminant states \cite{supplemental}. This suggests the importance of correlations and is the focus of future theoretical efforts. Nonetheless, independent of a microscopic model, a simple analysis supported by experimental observation suggests the time available for coherent manipulations is much shorter than the expected single-atom scattering time, placing significant constraints on Rydberg dressing proposals. Importantly, the mechanism described here scales unfavorably with principal quantum number \cite{supplemental} and implies the need to account for even a small number of impurity Rydberg atoms when considering interactions in dense gases. And although we have focused on exciting to an s-state with contaminant p-states, this mechanism is similarly present for excitation to any Rydberg state, which will populate nearby states of opposite parity. We note similar broadening has been observed in Rydberg transitions in strontium \cite{DeSalvo2015}.

\begin{acknowledgments} The authors thank T. C. Killian, F. B. Dunning, A. Browaeys, M. Foss-Feig, A. Hu, R. M. Wilson, and Z. X. Gong for helpful discussions. This work was partially supported by NSF PIF, AFOSR, ARO, ARL-CDQI, and NSF PFC at JQI.  
\end{acknowledgments}


%

\newpage
\setcounter{equation}{0}
\renewcommand{\theequation}{S\arabic{equation}}
\setcounter{figure}{0}
\renewcommand{\thefigure}{S\arabic{figure}}
\sloppy

\onecolumngrid
\begin{center}
\large{\textbf{Supplementary material: Anomalous broadening in driven dissipative Rydberg systems}}\\
\end{center}
\section*{}
\twocolumngrid

\section*{Rabi frequency calibration}
We calibrate the Rabi frequencies of the two optical fields using the light shifts due to each field. For the 795-nm laser, which is detuned $\Delta/2\pi\approx235$~MHz below the $\left|5s;F=2,m_F=-2\right\rangle\rightarrow\left|5p_{1/2};F=1,m_F=-1\right\rangle$ transition, we observe the intensity dependent shift, $\delta_{795}$, of the $\left|5s;F=2,m_F=-2\right\rangle\rightarrow\left|5s;F=1,m_F=-1\right\rangle$ microwave transition. The light shift $\delta_{795}=\Omega_1^2/4\Delta$ is a factor of $\Delta_{\mathrm{hyperfine}}/\Delta\approx30$ larger for the $\left|5s;F=2,m_F=-2\right\rangle$ state compared to the $\left|5s;F=1,m_F=-1\right\rangle$ state so we take the effect on the microwave transition as the light shift on the $\left|5s;F=2,m_F=-2\right\rangle$ state. We observe a shift that is linear in intensity $I$, as expected, and use this to calibrate the Rabi frequency $\Omega_1(I)=2\sqrt{\delta_{795}(I)\Delta}$ which we vary from $2\pi\times0.2$~MHz to $2\pi\times10$~MHz to control the two-photon Rabi frequency.

To obtain the Rabi frequency of the 485-nm field, detuned $\Delta/2\pi\approx235$~MHz above the $\left|5p_{1/2};F=1,m_F=-1\right\rangle\rightarrow\left|18s;F=2,m_F=-2\right\rangle$ transition, we observe the intensity dependent shift, $\delta_{485}$, of the two-photon $\left|5s;F=2,m_F=-2\right\rangle\rightarrow\left|18s;F=2,m_F=-2\right\rangle$ transition keeping the 795-nm light intensity (and frequency) constant. The 485-nm field is far from any transition coupling the ground $5s$ state to any optically excited state, so the shift of the $\left|18s;F=2,m_F=-2\right\rangle$ state is the dominant contribution and $\delta_{485}=\Omega_2^2/4\Delta$. We observe a shift that is linear in intensity, as expected, and use this to calibrate the Rabi frequency $\Omega_2(I)=2\sqrt{\delta_{485}(I)\Delta}$ which we hold at the maximum value given available laser power and beam diameter, $\Omega_2/2\pi\approx7$~MHz.

\section*{Lattice filling}
Loading $4\times10^4$ atoms into the 3D optical lattice as described in the main text leads to overfilling (more than one atom per site) in a fraction of the lattice sites. However, we determined that the filling fraction does not affect the broadening, which is only determined by the overall density. We made this determination by 
comparing a Mott insulator state with no more than one atom per site to a state with a Poissonian distribution of atoms per site, but the same total number of atoms in the participating ground state (and thus the same global density). Only the overall atom number was pertinent to the broadening. We also compared random transfer of half of the atoms to the participating ground state with transfer of all atoms on every other site in 2D in a checkerboard fashion \cite{Lee2007}. These two cases lead to a factor of two difference in the filling fraction per lattice site. We again saw that the total atom number controlled the broadening without regard to the microscopic configuration.

\section*{$18s$ Lifetime}
We observe 780-nm fluorescence on the $5p_{3/2}-5s$ transition following excitation to the 18s Rydberg state and fit a decaying exponential to extract a lifetime. The bulk of the fluorescence is due to the $\approx30~\%$ of the decay via $18s-5p_{3/2}-5s$ for which the 28-ns lifetime of the $5p_{3/2}$ is negligible compared to $\tau_0=3.5~\mu\mathrm{s}$, the natural lifetime of the 18s. However, $\approx1~\%$ of the population decays via channels of the form $18s-np-n's-5p_{3/2}-5s$ for which the lifetimes of the intermediate states delay the final 780-nm photon leading to effective lifetimes $\approx4\times$ longer \cite{Beterov2009} and an increase in the measured lifetime by $\approx10~\%$. In addition, radiation trapping, in which photons are reabsorbed and reemitted one or more times before leaving the cloud, can alter the measured lifetime. By measuring fluorescence following resonant excitation on the $5s_{1/2}-5p_{3/2}$ transition, we place a limit of $<0.5~\mu\mathrm{s}$, shorter than the measured lifetimes, on the characteristic time for radiation trapping, and thus expect little or no alteration of the measured lifetime due to this effect.

In all cases studied, including resonant excitation at different two-photon Rabi frequencies and detuned excitation at different detunings, the extracted lifetime is $8~\%$ to $20~\%$ longer than $\tau_0$, consistent with the natural lifetime and suggesting no shortening of the lifetime due to superradiance or other purely homogeneous effects. In addition, the amount of fluorescence is consistent with the optical pumping signal under all different conditions. This further suggests that superradiance is not a significant effect as it would lead to a Rydberg density dependence of the fractional decay via the $5p_{3/2}$ state compared to other intermediate states.

\section*{Dipole vs van der Waals scaling}
In order to calculate $\beta_3$, we have only included the $17p$ and $18p$ states and have excluded interactions that do not conserve total magnetic quantum number. The former should add $\approx1~\%$ to the total interaction and the latter are not resonant due to Zeeman splitting caused by a non-zero magnetic field. However some $m$ non-conserving interactions have an energy mismatch that is less than the largest observed linewidth and may play a role.

\begin{figure}[!h]
\includegraphics[width=6cm]{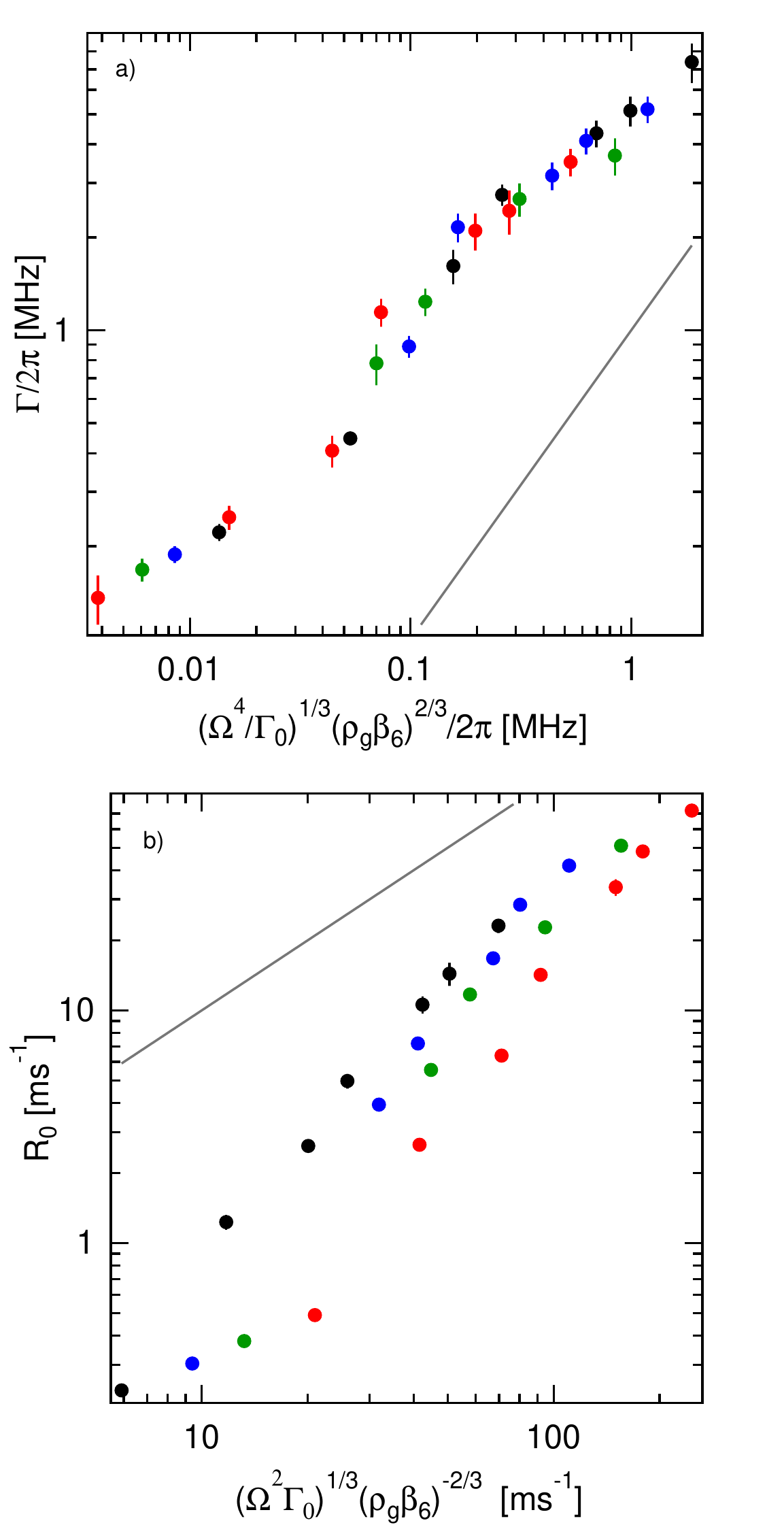}
\caption{(a) Measured $\Gamma$ and (b) $R_0$ for different $\Omega$ and $\rho_g$ vs expected scaling for van der Waals dominated interaction. Linear scalings with unit slope are indicated with solid lines.} \label{collapse6}
\end{figure}

If the broadening were due to van der Waals interactions between $18s$ atoms, one expects scaling $\Gamma=C_6 \rho_{18s}^2$. We define an interaction volume $\beta_6=\sqrt{C_6/\Gamma_0}$, in which case we assume $\Gamma=\beta_6^2 \rho_{g}^2R_0^2/\Gamma_0$. Combined with the observed relation $R_0 = \Omega^2/\Gamma$, we express $\Gamma$ and $R_0$ in terms of the independently controlled variables $\rho_g$ and $\Omega$: 
\begin{equation}
\begin{split}
&\Gamma=\left(\Omega^{4}/\Gamma_0\right)^{1/3} \ \left( \rho_g \beta_6\right)^{2/3}\\
&R_0=\left(\Omega^{2}\Gamma_0\right)^{1/3} \ \left( \rho_g \beta_6\right)^{-2/3}.
\end{split}
\end{equation}
Figure \ref{collapse6} shows $\Gamma$ and $R_0$ in terms of these expressions. There is a large mismatch in values between the data and the van der Waals scaling. In addition the data is not linear in the expressions and the rate $R_0$ does not collapse to a single function for all $\Omega$ and $\rho_g$. 

\section*{Dipole broadening in various regimes}
We consider how broadening due to dipole interaction with spontaneously created atoms in nearby Rydberg states affects a variety of possible operating regimes. For systems with small atom number, the time until the first atom is created in a nearby Rydberg state can be long compared to the lifetime if $bN_0\ll4\delta^2/\Omega^2$. However, since the dressed Rydberg interaction is of order $\Omega^4/\delta^3$ it may be difficult to access interesting many-body effects in this regime.

Additionally, this effect remains at higher principal quantum number: $\beta_3\propto n^{*7}$ where $n^*=n-\delta_{qd}$ is the effective principal quantum number and $\delta_{qd}$ is the quantum defect \cite{Li2003}. The contaminant density at which the dipole interaction overtakes the dressed interaction ($\Omega^2/\delta C_3$) scales as $n^{*-4}$. Assuming reasonable limits on attainable Rabi frequency ($\Omega/2\pi<100~\rm{MHz}$) and dressed interaction strength ($\Omega^4/\delta^3<2\pi\times10~\rm{kHz}$), this corresponds to a single spontaneous Rydberg atom in a typical 10-$\mu\rm{m}$ diameter ultracold atom system for $n>40$. Systems at higher principal quantum number may allow some additional time in an absolute sense due to their longer lifetimes (which scale as $<n^{*3}$), but are still subject to this effect and cannot access the expected single-atom scattering rates.

Since the van der Waals interaction is derived from the dipole interaction with nearby states, any configuration with large van der Waals interaction must have dipole interactions at least as large. Even in the case of operation at a F\"orster resonance to maximize the interaction \cite{pupillo2010,vanBijnen2015}, the uncontrolled, spontaneously created population in nearby Rydberg states causes interactions larger than any dressed interaction and must be considered in determining the timescales available for coherent operations. 

\section*{Inhomogeneous Gutzwiller Mean-Field Theory}

Treating each atom as a three level system with states $g$, $p$, and $s$, corresponding to ground, $18s$, and $np$ states respectively, we model the system with the following Hamiltonian:
\begin{equation}
H = - \delta \sum_i \sigma_i^{ss} + \Omega \sum_i (\sigma_i^{sg} + \sigma_i^{gs}) + \sum_{ij} V_{ij} \sigma_i^{sp} \sigma_j^{ps} + h.c.,
\end{equation}
where $V_{ij} = \frac{C_3}{r_{ij}^3} (1 - 3 \cos^2 \theta)$ is the dipole-dipole interaction with $\theta$ the angle between the $z$-axis and the relative position $\mathbf{r}_{ij}$. The system evolves according to the master equation
\begin{subequations}
\begin{equation}
\dot{\rho} = - i [H,\rho] + \mathcal{L}_s + \mathcal{L}_p + \mathcal{L}_R,
\end{equation}
\begin{equation}
\mathcal{L}_s = \Gamma_s \sum_i \left(\sigma_i^{gs} \rho \sigma_i^{sg} - \frac{1}{2}\{\sigma_i^{ss},\rho\}\right), 
\end{equation}
\begin{equation}
\mathcal{L}_p = \Gamma_p \sum_i \left(\sigma_i^{gp} \rho \sigma_i^{pg} - \frac{1}{2}\{\sigma_i^{pp},\rho\}\right), 
\end{equation}
\begin{equation}
\mathcal{L}_R = \Gamma_R \sum_i \left(\sigma_i^{ps} \rho \sigma_i^{sp} - \frac{1}{2}\{\sigma_i^{ss},\rho\}\right),
\end{equation}
\end{subequations}
where $\mathcal{L}_s$, $\mathcal{L}_p$, and $\mathcal{L}_R$ are Lindblad terms corresponding to decay from $s$ to $g$, from $p$ to $g$, and from $s$ to $p$ respectively.

Using an inhomogeneous Gutzwiller mean-field approximation, we assume the density matrix has the form
\begin{equation}
\rho = \bigotimes_i \rho_i,
\end{equation}
which assumes there are no correlations between different atoms. The method is inhomogeneous in the sense that each atom has its own density matrix, whereas in homogeneous Gutzwiller mean-field theory all atoms have the same density matrix. This results in an effective local Hamiltonian

\begin{equation}
H_i^{eff} = -\delta \sigma_i^{ss} + \frac{\Omega}{2} (\sigma_i^{sg} + \sigma_i^{gs}) + \sum_j V_{ij} \sigma_i^{sp} \langle \sigma_j^{ps} \rangle + h.c.
\end{equation}

In this picture, the interactions behave as an effective driving term between the $s$ and $p$ states whose strength and phase are determined by the $\langle \sigma^{ps} \rangle$ coherences of the surrounding atoms.

We determine the steady state of these equations numerically by initializing a cubic lattice of randomized density matrices for each site and evolving the system according to the master equation and effective Hamiltonian. This was done for a variety of $\Omega$ and $\delta$ based on the experimental measurements. The interaction strengths were reduced due to computational constraints, but the nearest neighbor interaction strength remained at least two orders of magnitude above $\Omega$ and all decay rates, compared to four in the experiment.

In all cases, we find that the $\langle \sigma^{ps}\rangle$ coherences all decay to zero in steady state, in which case the system behaves as if there are no interactions. This is a consequence of the flip-flop interactions and would not occur if the interactions were of the form
\begin{equation}
\sum_{ij} V_{ij} \sigma^{ss}_i \sigma^{pp}_j.
\end{equation}

Collective decay between the $s$ and $p$ states does result in nonzero $\langle \sigma^{ps} \rangle$ coherences in steady state, but the effect of interactions in this case is small and the experimental results indicate that collective decay is not the source of the observed broadening. 

Future work to solve this model using techniques beyond Gutzwiller mean-field will hopefully provide a more full understanding of the observed broadening and inform a more detailed study of the early time dynamics of the system.

\end{document}